\documentclass[journal]{IEEEtran}
\usepackage{cite}
\usepackage{amsmath}
\usepackage{array}
\usepackage{fixltx2e}
\usepackage{stfloats}
\usepackage{url}
\usepackage{times,amsmath,epsfig}
\usepackage{fancyhdr}
\usepackage{amsmath}
\usepackage{amsfonts}
\usepackage{amssymb}
\usepackage[latin1]{inputenc}
\usepackage{array}
\usepackage{graphicx}
\usepackage{url}
\usepackage{subfigure}
\usepackage{bm}
\usepackage{breqn}
\usepackage{xcolor}
\usepackage{soul}
\usepackage{amssymb}
\ifCLASSINFOpdf
\else
\fi

\begin{document}
\title{A Bore-Integrated Patch Antenna Array for \\Whole-Body Excitation in Ultra-High-Field Magnetic Resonance Imaging}

\author{Svetlana~S.~Egorova,
        Nikolai~A.~Lisachenko,
        Egor~I.~Kretov,
        Stanislav~B.~Glybovski,
        and~Georgiy~A.~Solomakha
\thanks{Svetlana~S.~Egorova, Nikolai~A.~Lisachenko, Egor~I.~Kretov, Stanislav~B.~Glybovski, and~Georgiy~A.~Solomakha are with School of Physics
and Engineering, ITMO University, St. Petersburg 197101, Russia (e-mail: s.glybovski@metalab.ifmo.ru). \textit{Corresponding author: Stanislav~B.~Glybovski}}}
\markboth{IEEE Transactions on Biomedical Engineering,~Vol.~XX, No.~XX, August~2024}%
{Shell \MakeLowercase{\textit{et al.}}: Bare Demo of IEEEtran.cls for IEEE Journals}
\maketitle

\begin{abstract}
Objective: To develop and evaluate a bore-integrated patch antenna array designed for whole-body excitation in ultra-high-field (UHF) magnetic resonance imaging (MRI) with improved transmit efficiency and address the limitations of existing RF coil designs. Methods: The proposed patch antenna array utilizes the MRI bore's RF shield as a functional component to enhance the RF magnetic field ($B_1^+$) distribution. Numerical simulations were conducted to compare the performance of the patch antenna array to bore-integrated stripline and local dipole arrays. A decoupling structure was implemented to minimize coupling between adjacent patch antennas. The performance of the patch array was evaluated experimentally. Results: The proposed patch array provides 3.9 times higher averaged transmit (Tx) efficiency in the CP mode and 3.0 times higher for the phase shimming regime versus the bore-integrated stripline array. Conclusion: Compared to the stripline array, the bore-integrated patch antenna array offers significant improvements in Tx efficiency for whole-body UHF MRI. Significance: The findings support the feasibility of integrating arrays into the RF shield of MRI scanners. This could broaden the clinical use of UHF body MRI technology.
\end{abstract}

\IEEEpeerreviewmaketitle

\section{Introduction}

\IEEEPARstart{M}{}agnetic-resonance imaging (MRI) is a powerful method for high-quality and non-invasive investigation of the human body for clinical and research purposes. Currently, there is a clear trend towards increasing the static magnetic field ($B_0$ aligned in parallel to the bore axis $z$) of a superconductive magnet, improving the signal-to-noise ratio (SNR)\cite{6778087,ladd2018pros}. Recently, the first ultra-high-field (UHF, $B_0>7$~T) MRI systems were approved for limited clinical use for head and extremity regions \cite{FDA_7t}. However, some obstacles prevent this technology from being approved for the rest of the human body. A major limitation of 7T body imaging is the strong inhomogeneity of the transmit RF magnetic field ($B_1^+$) caused by interference patterns at relatively short RF field wavelengths at 298 MHz of around 12 cm (for brain tissues) \cite{erturk2019evolution}. This so-called \textit{dielectric effect} makes the $B_1^{+}$ distribution created by any single or double port volumetric coil significantly inhomogeneous either in a human head or a torso producing dark voids and deteriorating SNR \cite{vaughan20017tvs4t,7Tesla_Body}. The effect is especially harmful for cardiac, liver, and pelvis MRI scans.

To overcome this limitation, the parallel transmission (pTx) method (or \textit{RF shimming}) was proposed \cite{pTx} and developed for improving whole-body \cite{van2005b1,vaughan20069} imaging. This approach allows for homogenizing or focusing the transmit field distribution in an ROI by using multiple individually driven channels with customized phases and/or magnitudes of signals feeding an array of antenna elements mounted on the subject or surrounding it\cite{metzger2008local}. An additional benefit of the RF shimming approach is using multiple transmitters instead of just one to increase the total deposited RF power (within applicable SAR limitations) to increase the $B_1^+$ in ROI and image quality. In the subsequent works, more advanced and complicated methods combining modulation of RF and gradient magnetic fields of pTx like kT-points\cite{cloos2012kt} were developed. To maximize the ratio of the $B_1^{+}$ level at a given total accepted transmit power, referred to as \textit{transmit (Tx) efficiency}, array antenna elements are typically placed as close as possible to the body (in a local configuration) with appropriate dielectric spacers to control SAR value\cite{Alex_MRM,erturk201716}. However, this approach is currently used only for research purposes in volunteer studies because of the reduced comfort and safety issues caused by high local peak SAR. Despite the possibility of SAR management using RF shimming, this procedure is currently not acceptable for clinical applications. One of the ways to mitigate the comfort and SAR limitations for body imaging while keeping the capabilities of $B_1^{+}$ shimming and creating a clinic-like workflow is to place array elements away from the patient \cite{deniz2016radiofrequency}. In \cite{pavska2018rigid}, the Tx dipole elements were mounted on a special 3D-printed housing, providing more free space than typical local array configurations. Further increasing the array diameter leads to array designs with low-profile antenna elements conveniently integrated between the bore lining and gradient system. This leaves even more space for a patient and dedicated local receive (Rx) array. The corresponding approach is called \textit{bore-integrated array coils}.

In comparison with 1.5T, where no RF shimming is employed for body imaging, in 3T clinical scanners (128 MHz) bore-integrated birdcage coils \cite{birdcage} can be driven using two independent channels with orthogonal linear polarization \cite{brink2015clinical,murbach2016virtual}. However, birdcage coils may become radiating and hard to tune at higher frequencies as their circumference and longitudinal length become electrically large \cite{TEM}. Currently, there are no 7 T full-body scanners that have bore-integrated body coils. The first feasibility results of whole-body imaging at 7T were obtained using a TEM body coil composed of parallel longitudinal transmission striplines capacitively connected to the shield and to each other at their terminations \cite{7Tesla_Body}. However, it was concluded that the coil had limited performance in CP mode as the regions of interference minima could not be imaged within applicable SAR regulations. To have the ability to control $B_1^+$ via different pTx methods, one should use multiple amplifiers connected to several decoupled antenna array elements. Therefore, an important research direction is finding power-efficient multichannel array designs operating near a metal RF shield (bore). A bore is required to minimize the RF coil's interaction with other MR scanner systems. In this way, several approaches have been introduced, including highly decoupled stripline antennas \cite{orzada2017analysis,orzada201932,fiedler2021performance}, traveling-wave MRI \cite{zhang2012whole}, and high-pass inductively-coupled loop resonators \cite{gokyar2021electrically}.

In \cite{orzada2017analysis}, an eight-channel array of 20-mm-thick grounded stripline antenna elements was integrated into the space between the inner cladding of the magnet (bore liner) and the gradient coil of a 7T system. These elements were optimized for local configuration but applied in the integrated array due to their low coupling without further optimization. The comparison between the bore-integrated and local arrays with the same type and number of elements showed relatively low Tx efficiency of the integrated array. 
To solve this problem, it was proposed to increase the number of antenna elements (including those placed along the z-axis). This concept was studied further in \cite{fiedler2021performance}. The main goal was to increase the total power while simultaneously lowering the power per element and increasing the degrees of freedom during RF shimming and pTx. The latter leads to improved control over the field uniformity. The most significant improvement was observed in the step from 8 to 12 elements in one row. Additionally, adding more rows further increased the longitudinal coverage. Finally, a 32-channel system based on stripline antennas showed promising performance for UHF whole-body imaging \cite{orzada201932}. The low transmit efficiency was partly compensated by the 32 kW total peak power provided by the custom-built amplifiers. However, it was stated that a more efficient design for the body array antenna elements could lead to further improvements.

In this work, we demonstrate the concept of a novel bore-integrated patch array with improved efficiency compared to previously used stripline antennas (as in \cite{orzada2017analysis,orzada201932,fiedler2021performance}). Such stripline antenna elements have relatively low Tx efficiency in the bore-integrated configuration for the whole body. This can be explained by the poor radiation efficiency of a narrow shielded metal strip with current caused by the destructive interference effect. Note that low Tx efficiency means that a high fraction of accepted power is dissipated inside the antenna element instead of being radiated and then absorbed inside the human body. The efficiency of a stripline element can be improved by using a high-impedance surface as a ground plane \cite{chen2016electromagnetic}, which is too complicated to implement for an entire whole-body MRI bore. Instead, we propose using wide-patch antennas with two relatively long radiating slots with much higher radiation efficiency as integrated array elements.
\textit{Patch} (or \textit{microstrip}) antennas were first reported in the 1950s and, since the 1970s, have been widely used in radar, communication, and navigation systems \cite{pozar1992microstrip}. The design principles of patch antennas can be found, e.g., in handbooks \cite{bahl1980microstrip,balanis2015antenna}, or review papers \cite{1142523,pozar1992microstrip}. Despite patch antennas have been previously proposed as local transmit elements for MRI oriented in the coronal (for human body imaging at 7T \cite{webb2010mri}, and for small-animal imaging at 14 T \cite{gandji2018development}) or transverse (for human brain imaging at 9.4 T \cite{hoffmann2013human}, 10.5 T \cite{bluem2018excitation}, and small-animal imaging at 16.4 T \cite{shajan2012rat}) plane for $B_1^{+}$ homogenization, their great potential of power-efficient radiation in the presence of a metal shield was never applied to the authors knowledge to bore-integrated configurations. Here, we compare the operation of the patch element and the stripline one by evaluating two corresponding eight-element arrays numerically (by simulations on a homogeneous phantom mimicking a human torso and a multi-tissue voxel model of an adult male) as well as experimentally (on the bench). A drastically higher transmit- and SAR efficiency is obtained using patch elements. Moreover, the methods for tuning, matching, and decoupling the patch elements specific to the bore environment are proposed and discussed. A preliminary version of this work has been reported in \cite{egorova2023numerical}.

\section{Methods}
\subsection{A bore-integrated patch antenna as an array element}

The idea behind the proposed bore-integrated array is as follows: once bore-integrated antenna elements are phased in a CP mode (with a phase step of -45$^{\circ}$ in an eight-channel configuration), their $B_1^+$  fields interfere constructively inside the subject. At the same time, the radiative power leakage outside of the bore is suppressed, allowing the subject to absorb significant power. Consequently, the Tx efficiency of the integrated array can be improved by maximizing the radiation efficiency of each single antenna element in free space. For the shielded case, this task can be solved by choosing patch radiators with resonant lengths along $z$ and, as long as possible, radiative slots in the transverse direction.

\begin{figure*}
\center
\begin{minipage}{1\linewidth}
\center{\includegraphics[width=0.85\linewidth]{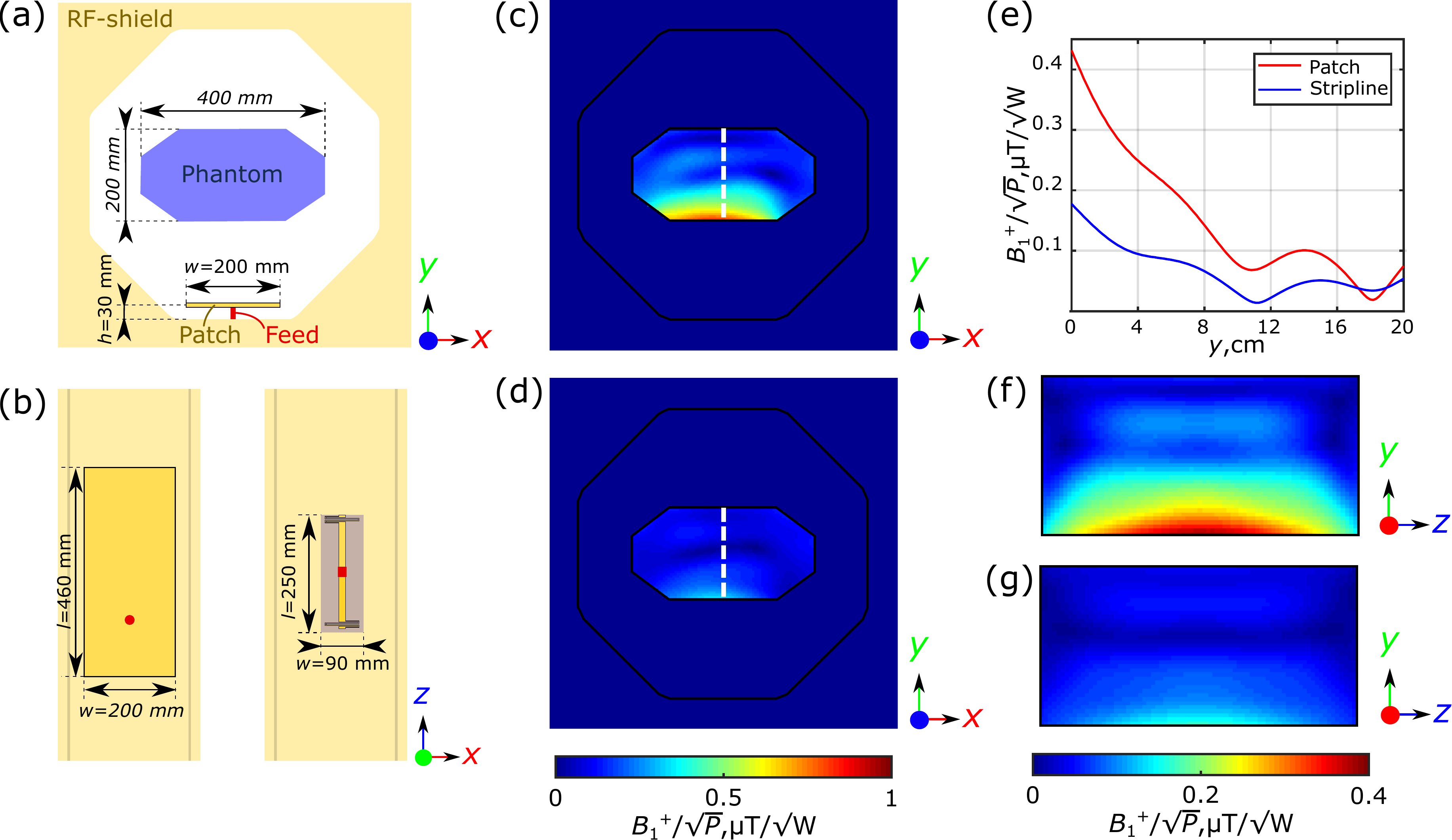}}
\end{minipage}
\caption{The geometry of the proposed bore-integrated patch antenna and results of its numerical comparison with the stripline antenna \cite{orzada2017analysis}: (a) position of one patch antenna element vs. the octagonal RF shield and a homogeneous phantom; (b) geometry of the patch and stripline antenna elements (top view); distributions of Tx efficiency in the central transverse slice of the phantom for the patch (c) and stripline (d); (e) in-depth profiles (from bottom to top) of the Tx efficiency along the dashed line indicated in panels (c,d); distributions of Tx efficiency field in the sagittal plane for the patch (f) and stripline (g).}
\label{fig1}
\end{figure*}

As a base for an integrated configuration of antenna array mounted on an RF shield with a diameter of 675 mm (same as in \cite{orzada2017analysis}). The corresponding inner surface of the shield serves as a common ground plane for eight identical rectangular patch antennas, each consisting of a flat rectangular metal plate with width $w$ and length $l$. The geometry of one antenna in the transverse and coronal planes is illustrated in panels (a) and (b) of Fig. \ref{fig1}, accordingly. 

For design and fabrication simplicity, we consider an octagonal profile of the bore with a 675 mm largest vertex-to-vertex distance, as shown in Fig.~\ref{fig1}(a). Each flat section of this bore is a ground plane with one of eight flat rectangular patches placed at a distance $h$ from it. Alternatively, one could employ cylindrical circular patch antennas of the same thickness $h$ following the curvature of a cylindrical RF shield\cite{krowne1983cylindrical,song2018systematic}. 

For conventional patch antennas, $h$ is typically chosen from a range of $0.003\lambda_0  \leq h \leq  0.1\lambda_0$ with $\lambda_0$ as the wavelength in free space. Larger $h$ provides wider bandwidth and lower dissipative losses. However, in this work, $h=30$ mm is chosen to fit in the space between the bore liner and the gradient coil, similarly to \cite{orzada2017analysis}. Note that thicker patch antennas could be preferable for MR scanners with wider bores. The space between the patch and the ground plane is filled with foam with a dielectric permittivity close to one (so-called air substrate) to minimize losses. In this case, like for conventional patches, $ l\lesssim \lambda_0/2$, which tunes the patch resonator to the resonant frequency of TM$_{010}$ eigenmode at the Larmor frequency of 298 MHz. In this regime, the antenna's radiation can be described by two in-phase slot radiators of width $w$ separated by distance $l$, creating maximum radiation towards the bore center. The value of $w$ should also be maximized to minimize losses. However, for a given bore circumference and $N=8$ array elements, width $w$ is limited by mutual coupling between adjacent antennas. For the given configuration, the resonant patch length is $l=446$ mm, while $w=200$ mm has been set to leave a gap no narrower than $h$ between similar patches in the array configuration. 

The proposed patch antenna is fed conventionally with a wire probe connected to a coaxial cable through the hole in the ground plane. The feeding point is shifted 51 mm off the patch center in the $z$-direction to match the input impedance to the line impedance of 50 Ohm. Additionally, a lumped trimmer series capacitor is used at the feeding point to compensate for a parasitic feed inductance.
For comparison, the stripline antenna element reproducing the geometry used in \cite{orzada2017analysis} mounted on the same face of the octagonal RF shield (see left sub-panel of Fig. \ref{fig1}(b)) is considered. Both antennas are compared by studying their $B_1^{+}$ distributions created in a pelvis-shaped phantom (mimicking the human body at 7T with homogeneous dielectric properties $\varepsilon_{\text{r}}=34$, $\sigma=0.45$ S/m) shown in Fig. \ref{fig1}(a) at the same power at their matched inputs. The calculations were made using the finite-element method in the frequency domain (FEM-FD) implemented in CST Studio Suite 2021 (Dassault Systemes, Velizy-Villacoublay, France).

\subsection{Decoupling of two adjacent patch array antennas}

An eighth-element array of patches with maximized width $w$ suffers from high mutual coupling. Although H-plane coupling between adjacent patches in conventional antenna arrays radiating to free space is relatively low, it becomes an issue for the considered densely packed patches in the bore environment. It was previously shown that patch antennas positioned nearby on a concave metal shield are coupled drastically stronger than the same patches on a planar or convex shield having the same separation \cite{6599841}. Our simulations show that the transmission coefficient $|S_{21}|$ between matched inputs of two patches is also a dependent of $w$. Thus, for the considered inter-element separation inside the bore, $|S_{21}|$ takes values from -12 dB to -3 dB depending on $w$. Eventually, width maximization for the efficiency of patch elements in the eight-element integrated array configuration requires additional decoupling.

Several decoupling techniques have been proposed for patch antennas with small ($<\lambda_0/10$) H-plane separation. Among them, there are the following: insertion of an asymmetrical coplanar strip wall \cite{qi2015mutual} (a combination of a metal wall with a passive dipole antenna), insertion of narrow interdigital lines \cite{qi2015improving}, and common-differential mode cancellation \cite{sun2020decoupling}. However, to decouple patches placed on a concave shield in the absence of a common dielectric substrate and to be able to adjust the decoupling during the experiment finely, we proposed a new decoupling structure. 

\begin{figure}
\center
\begin{minipage}{1\linewidth}
\center{\includegraphics[width=0.65\linewidth]{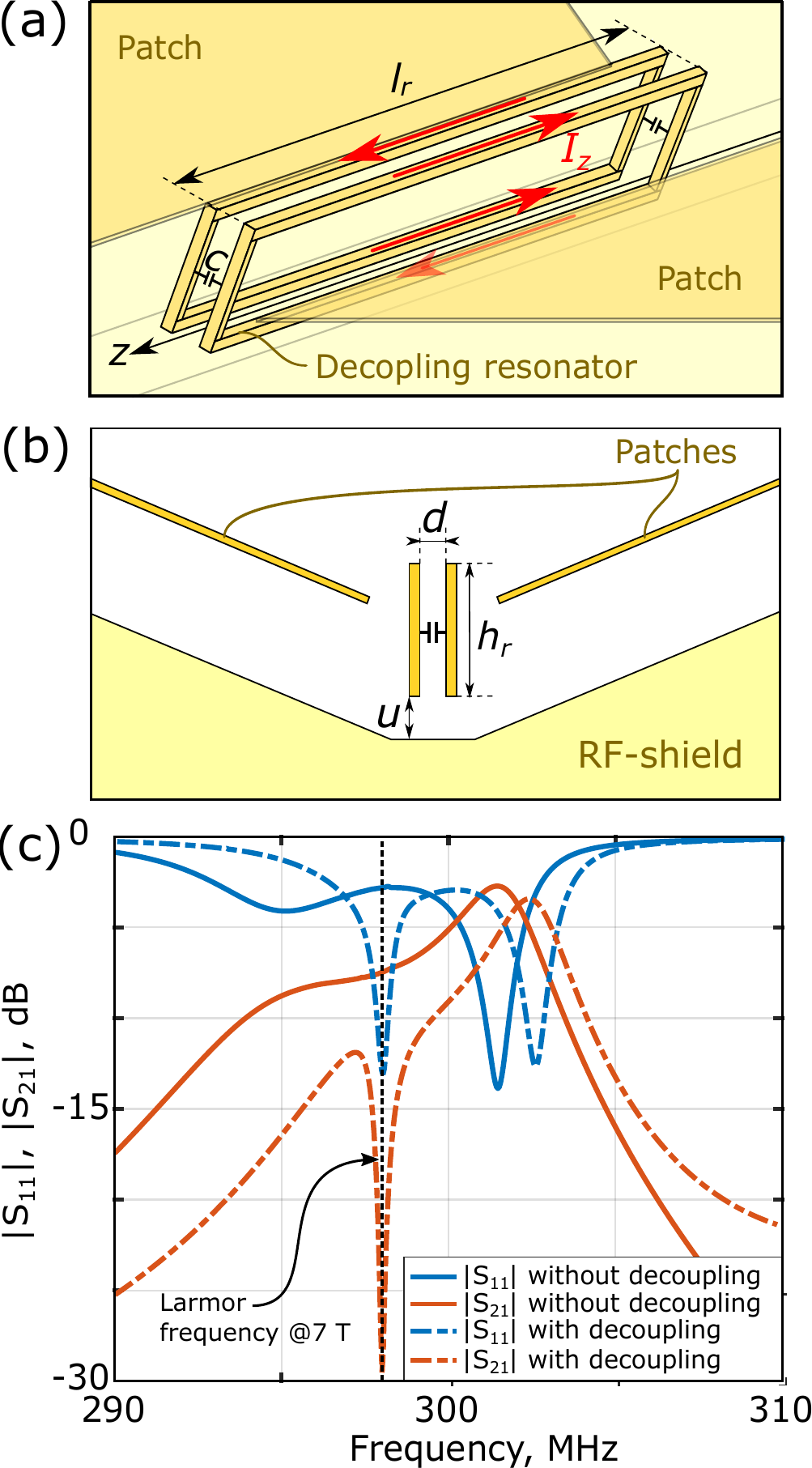}}
\end{minipage}
\caption{Geometry and results of a parametric numerical study of the proposed decoupling structure: (a) general view of the dual-loop decoupling resonator (arrows show the directions of longitudinal currents related to the differential coupled eigenmode); (b) geometric parameters of the structure; (c) numerically calculated $S$-parameters for two bore-integrated patch antennas with and without the decoupling structure.}
\label{fig2}
\end{figure}

The proposed structure is implemented as a passive arrangement of two rectangular metal loops of length $l_r <  \lambda_0/2$ and height $h_r\approx h$ placed in the gap between the patches as shown in Fig. \ref{fig2}(a,b). 
The planes of two loops have separation $d$, while both loops are elevated over the concave RF shield at a height $u$ as shown in Fig. \ref{fig2}(b). Both loops support an eigenmode for which antiparallel currents flow along two longitudinal conductors (parallel to $z$ axis). Since two loops act as two highly-coupled resonators, the structure supports their differential coupled mode (DM), which has a configuration of the induced currents shown in Fig.~\ref{fig2}(a). By exciting the DM, the additional coupling between the patches obtains an opposite sign from the initial (direct) coupling. The decoupling condition is achieved by adjusting the DM's resonant frequency by changing the tuning capacitors $C$ connecting the two loops as shown in Fig.~\ref{fig2}(a), and by adjusting the loop area. The latter is achieved by elongating four longitudinal conductors of the decoupling structure in a way similar to telescopic dipoles. 
The above-discussed geometric parameters were determined through parametric numerical simulations in CST Microwave Studio 2021, where two patch antennas and the decoupling structure were simulated inside the bore in the presence of the same phantom as in the previous subsection.
While $d$, $h_r$, and $d$ were chosen to fit in the gap between the patches and fixed, parameters $l_r$ (affecting both the DM's frequency and its excitation amplitude) and $C$ (affecting only the DM's frequency) were adjusted during the simulations to get position the narrow minimum of $|S_{21}|$ at 298 MHz. The S-parameters of two adjacent patch antennas before and after decoupling are compared in Fig.\ref{fig2} (c).

\subsection{Eight-element phased array}

After tuning, matching, and decoupling two adjacent patch antennas as described in the previous subsections, an eight-element bore-integrated phased array composed of similar patches and optimized decoupling structures was evaluated numerically. Two eight-element arrays of patch and stripline antennas are compared for the same RF shield size and phantom as above. 
Capacitance, $C$ of the connecting capacitors in the decoupling structures and their length $L_r$, are adjusted to achieve a decoupling of better than 15 dB between the inputs of the matched patches. Note that the striplines, when matching using a matching circuit from \cite{orzada2017analysis}, show sufficient decoupling of better than 15 dB.
Following the approach of the works, \cite{orzada2017analysis,orzada201932}, the circularly-polarized (CP) mode was driven by setting a progressive phase shift of -45$^{\circ}$ to the input of subsequent antennas around the circumference of the array. The calculated $B_1^{+}$ maps are normalized to the square root of total accepted power to calculate Tx efficiency.

\subsection{Experimental study}

\begin{figure}
\center
\begin{minipage}{1\linewidth}
\center{\includegraphics[width=0.85\linewidth]{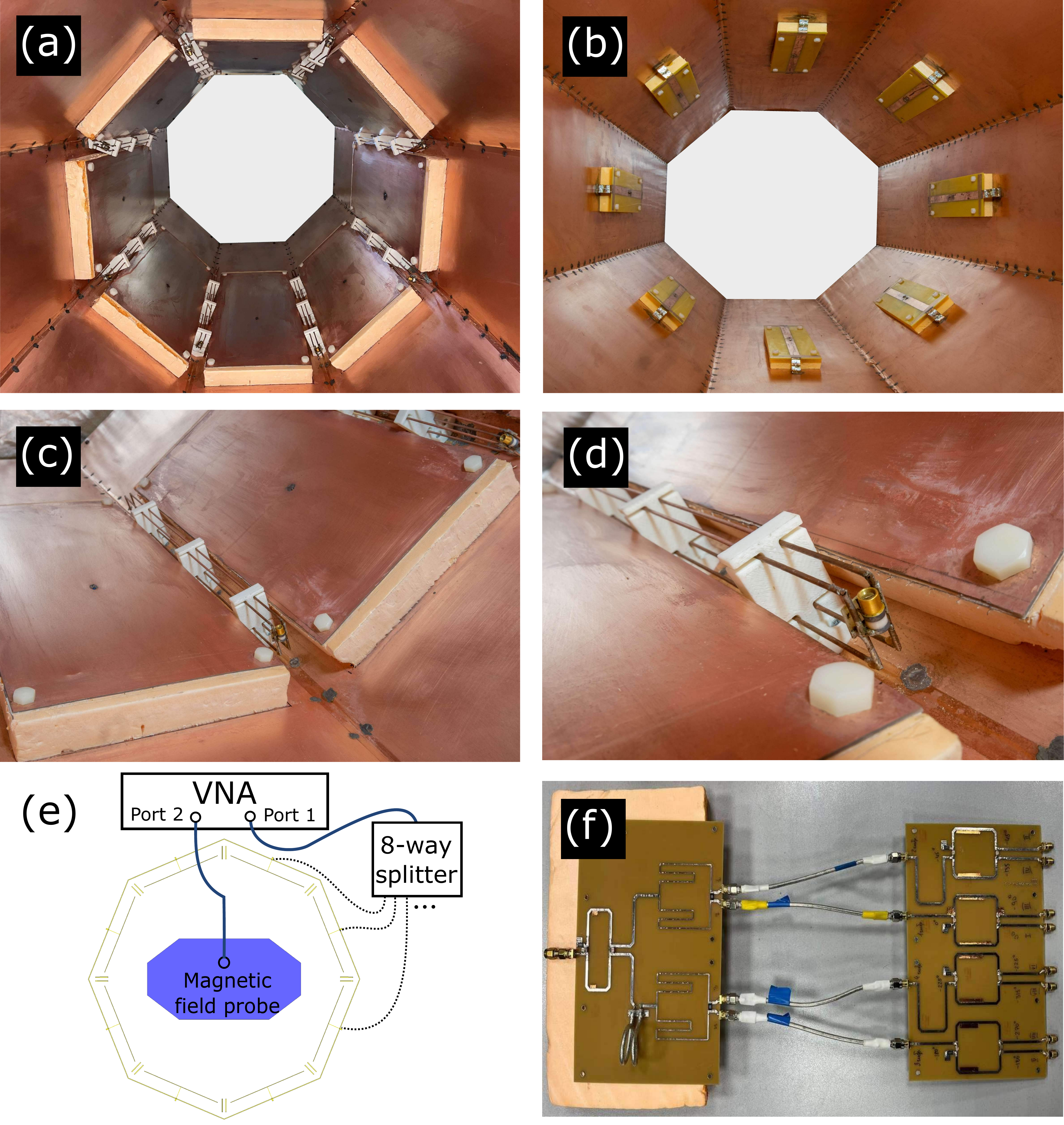}}
\end{minipage}
\caption{Experimental setups for $B_1^+$ field measurements in a CP mode excitation regime: (a) patch array prototype; (b) stripline array prototype; (c,d) adjacent patch antennas with a decoupling structure; (e) schematic of the near-field measurement setup; (f) photograph of the 8-channel microstrip splitter used for CP excitation.}
\label{fig3}
\end{figure}

An experimental full-sized prototype was built in our lab to confirm the numerical simulation results and verify the improved efficiency of the proposed patch array over the stripline array. Both arrays had an octagonal RF shield of the same geometry as in the numerical simulation composed of eight copper-cladded 1.5-mm-thick FR4 sheets soldered together along contacting edges. The entire shield was mounted on a rigid wooden frame. The patch antennas were made of similar copper-cladded 1.5-mm-thick FR4 sheets fixed to the RF shield with 30-mm-thick foam brick acting as antenna substrate. The photo of the patch array is shown in Fig.~\ref{fig3}(a), while two of eight patches with a decoupling structure are shown in detail in Fig.~\ref{fig3}(c,d). 
The decoupling loops are made of 2-mm-diameter copper wire and connected by a circuit board with a soldered variable capacitor (1-19 pF, Johanson, Boonton, NJ, USA) (see Fig.~\ref{fig3}(d)). Before measurements, the length of the loops and the capacitance are adjusted to minimize coupling between adjacent patches. The probe excitation method of the patch antennas is used with a probe position determined during the simulations, compensating the parasitic probe inductance by adding a series capacitor at the feeding point. With this aim, a trimmer capacitor (1-19 pF, Johanson, Boonton, NJ, USA)  is used at each input, soldered to a small circuit board placed outside the RF shield. The stripline array built to compare the efficiency is shown in Fig.~\ref{fig3}(b). The design of the stripline array element, balun, and matching network is similar to that was presented in work\cite{orzada201932}. 

During the RF field measurements, both arrays are driven in the CP mode utilizing a self-made 8-channel microstrip splitter, which input is connected to the first port of a vector network analyzer (VNA) TR1300/1 (Copper Mountain Tech, Indianapolis, USA)  schematically shown in Fig.~\ref{fig3}(e). The splitter shown in Fig.~\ref{fig3}(f) consists consisted of five 90$^{\circ}$ hybrid couplers, two Wilkinson power splitters, two transmission line -45$^{\circ}$ phase shifters and one  -90$^{\circ}$ phase shifter. The splitter outputs are connected to the inputs of array elements using low-loss RG-58 coaxial cables with lengths of 2 m. The same splitter and cables are used to characterize the stripline array. The RF magnetic field created by both arrays to be compared is measured using a near-field scanner with a 3D stepped positioning of a magnetic probe connected to the second port of the same VNA (see Fig. \ref{fig3}(e)). A 14-mm-diameter XF-R 100-1 (Langer EMV-Technik GmbH, Bannewitz, Germany) near-field magnetic field probe, mounted to the wooden arm rigidly connected to the 3D positioning system using a 3D printed interface plate is used.
The probe can freely move inside the phantom filled with a special solution consisting of distilled water, isopropyl alcohol, and table salt (NaCl), having the same permittivity and conductivity as a phantom in simulations.
The liquid is poured into a polycarbonate container with a wall thickness of 2 mm and the same shape and size as the phantom considered in the numerical simulations. The permittivity and conductivity of the phantom solution are measured using a SPEAG DAK-12 (SPEAG, Zurich, Switzerland) precision dielectric probe kit. 

The $B_1^+$ field distributions of the patch and stripline arrays are measured in the central transverse slice in the rectangular region with sizes $110\times 80$ mm$^2$ with a resolution of 5 mm. The output power of the VNA is set to 3 dBm. Since magnetic field probes are used, the spatial distribution of $S_{21}$ could be interpreted as a map of the complex amplitude of the corresponding Cartesian $B$-field component. Both $x$ and $y$ components of the $B$-field are measured during consequent scans (with different probe orientations) and then combined at the post-processing step to find the $B_1^+$ at each measured spatial position using MatLab 2020 software (The MathWorks, Inc., Natick, Massachusetts, USA). 

\subsection{Numerical simulations with models of the human body}

To evaluate the specific absorption rate (SAR) and $B_{1}^{+}$ distributions created inside the human body, the finite-integration technique in the time domain (FIT-TD) of CST Microwave Studio is used with a Duke voxel model (Zurich MedTech, Zurich, Switzerland). The proposed patch array is compared for the same number of elements to the bore-integrated stripline array\cite{orzada2017analysis} as well as the \textit{state-of-the-art} local (on-body) array of fractionated dipoles\cite{Alex_MRM}. Each model contains approximately $70\cdot10^{6}$ mesh cells. Local mesh settings on the surface of the array elements are applied to improve the accuracy of simulations. All array elements are ensured to be tuned, matched to 50 Ohm, and pairwise decoupled with a level of at least -10 dB. The prostate is chosen as a region of interest (ROI) for the $B_{1}^{+}$ comparison since it is located deeply in the human body like it was done in  \cite{metzger2008local,raaijmakers2016fractionated,solomakha2021self}. 

Fields created by individual channels for the patch and stripline arrays are combined to form CP mode. Additional phase-only RF shimming\cite{metzger2008local} is performed to maximize the field in the prostate region for the local dipole array. The obtained $B_1^+$ distributions are normalized to the square root of the accepted power $P$ to calculate each array's transmit (Tx) efficiency. In each case, SAR$_{10g}$ is calculated inside the entire voxel model, and the local peak value  pSAR$_{10g}$ is found. CST Legacy SAR calculation method is used. Averaging is performed over the 10g of tissues. For the comparison means, SAR-efficiency (Tx efficiency in the ROI normalized to the $\sqrt{\textnormal{pSAR$_{10g}$}}$ is calculated. 

\section{RESULTS}

\subsection{Comparative simulation of single antenna elements in the presence of a homogeneous phantom}

Numerically calculated $B_1^+$ distributions normalized to $\sqrt{P}$ (Tx efficiency) in the central transverse slice of the phantom, where $P$ is the accepted power, are presented Fig.~\ref{fig1}(c) and Fig.~\ref{fig1}(d) for the single patch antenna the stripline, accordingly. The penetration profiles of Tx-efficiency (from bottom to top of the phantom along the white dashed lines displayed in Fig.~\ref{fig1}(c,d)) are compared in Fig.~\ref{fig1}(e) for the same antenna elements. Finally, the distributions on the central sagittal plane are shown in Fig.~1(f) and Fig.~1(g) for the patch and the stripline, accordingly. From the results, one can conclude that the patch antenna element creates a considerably higher $B_1^+$ level inside the phantom than the stripline at the same accepted power while providing the same longitudinal coverage. Thus, the transmit efficiency is observed to be 2.45 times higher at the nearest face of the phantom and 2.22 times higher at a depth of 80 mm inside the phantom (which corresponds to a typical depth of the prostate inside the pelvis). This benefit in the transmit efficiency is observed even for single antenna elements (without CP mode excitation of an array). It can be attributed to lower dissipation losses inside the patch than the stripline (83 \% vs. 11 \% of the accepted power is dissipated inside the antenna element).

\subsection{Decoupling of two adjacent patch antennas in the MRI bore environment}

Our numerical simulations show that two proposed patch antennas exhibit high mutual coupling. The level of $S_{21}$ for two such antennas initially matched in single-element configuration reaches -5 dB as shown with the blue solid curve in Fig.~\ref{fig2}(c). Note that $|S_{21}|$ spectrum has two distinct minima corresponding to the resonances of two coupled modes of the patches (one above and one below the Larmor frequency).

\begin{figure}
\center
\center{\includegraphics[width=0.75\linewidth]{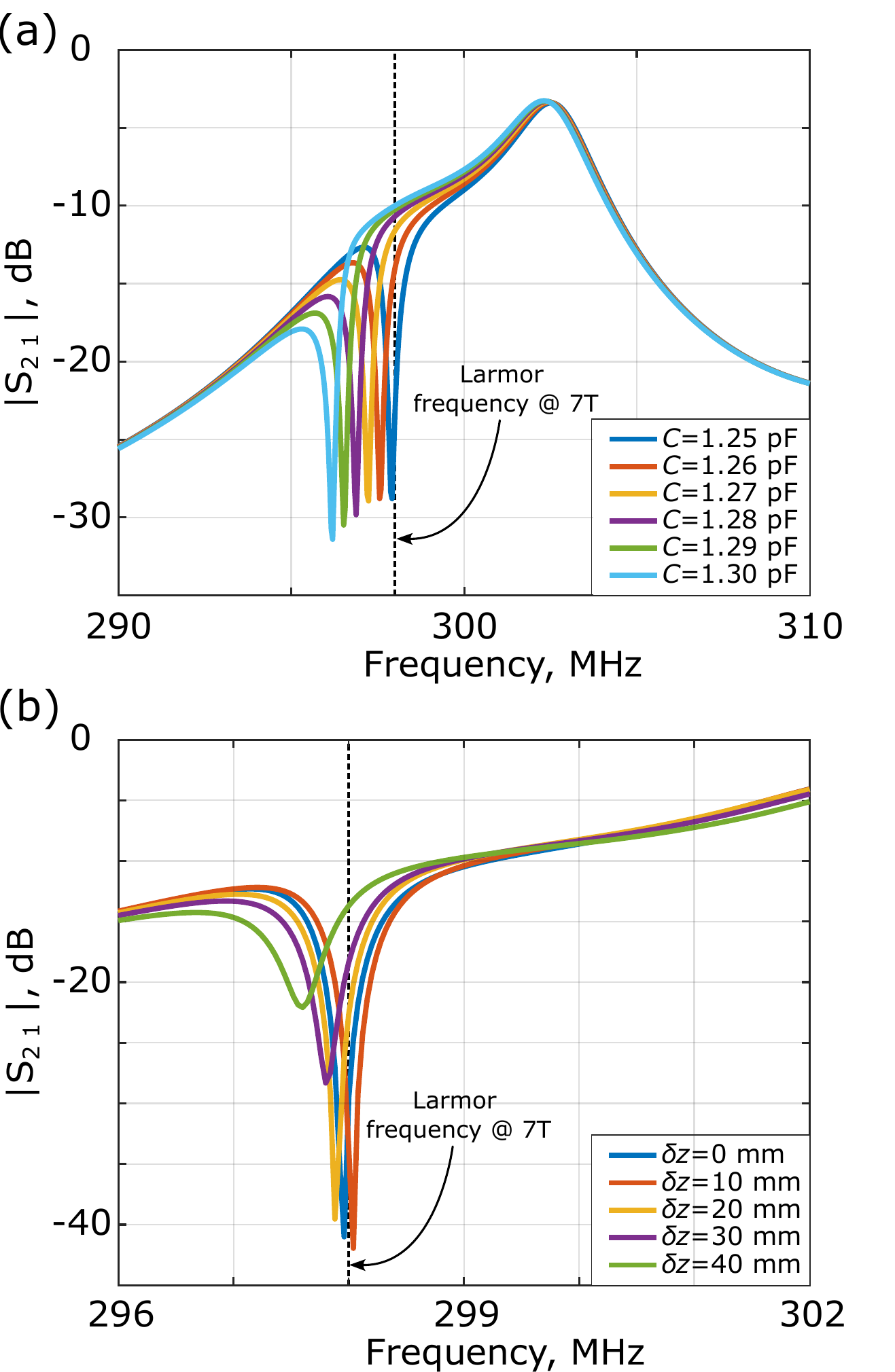}}
\caption{Numerical illustration of the possibility to tune the decoupling frequency (a) and the level of $|S_{21}|$ in its minimum (b) by adjusting capacitance $C$ and longitudinal shift $\delta z$ (b) of the decoupling structure.}
\label{fig4}
\end{figure}

Inserting the proposed decoupling structure allows one to significantly reduce $|S_{21}|$ in a narrow resonance range of the DM. During numerical simulations, it was found that the decoupling frequency can be adjusted by changing either $C$ or $l_r$ via DM mode tuning. For initial numerical optimization, the length is preferable, while for fine adjustment in the experiment, the capacitance parameter should be chosen for convenience. At the same time, the level of $|S_{21}|$ at the DM resonance is determined by the excitation amplitude of that mode and can be initially set in the simulations by varying $h_r$, while the lateral shift $\delta z$ of the resonator relative to the common H-plane symmetry plane of both patches is recommended for fine adjustment in the experiment. The effect both $C$ and $\delta z$ on $|S_{21}|$  is illustrated in Fig.~\ref{fig4}.

The optimization of the decoupling resonator resulted in the following parameters: $l_r=400$ mm, $h_r=40$ mm, $C = 1.25$ pF, $u = 20$ mm, $d = 6$ mm, and $\delta z=0$. After inserting the optimized resonator, the  $S_{21}$ reduces to -30 dB at 298 MHz, where matching is to be corrected to reach $|S_{11}|\approx-13$ dB via a small lateral shift of the feeding point in comparison to the single-element simulation. Note that better matching could be obtained using additional matching circuit elements connected to the inputs of both patches. The final numerical S-parameters after decoupling are shown in Fig.~\ref{fig2}(c) with dashed lines.

\subsection{The comparison of eight-channel arrays loaded with a phantom}

\begin{figure}
\center
\begin{minipage}{1\linewidth}
\center{\includegraphics[width=0.6\linewidth]{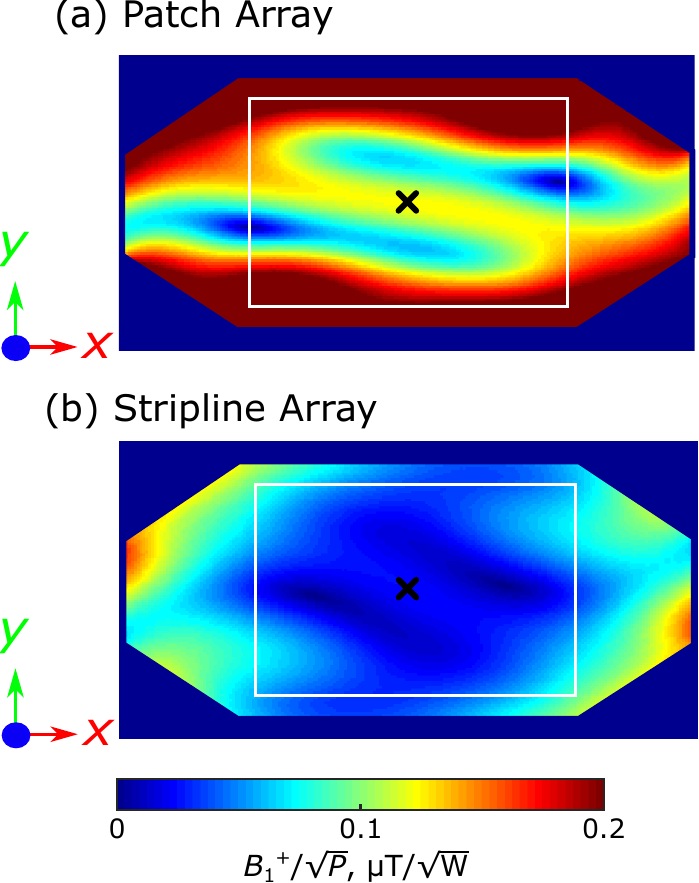}}
\end{minipage}
\caption{Numerically calculated distributions of the Tx efficiency field in the central transverse plane of the homogeneous phantom created by eight-channel arrays composed of patch (a) and stripline (b) antennas. The rectangles show the region of experimental comparison. The crosses show the center of the phantom.}
\label{fig5}
\end{figure}

To compare the Tx efficiency of two eight-element arrays, i.e., of patch and stripline antenna elements, we present their  $B_1^{+}$ distributions in the homogeneous phantom normalized to the total applied power obtained numerically and experimentally. Thus, the numerically calculated $B_1^+$ maps in the central transverse slice of the phantom for both arrays are presented in Fig.~\ref{fig5}. Note that the shapes of both maps (including the positions of zeros and lobes) are similar and specific for the considered phantom because of the CP mode excitation. At the center of the phantom marked with a cross in Fig.~\ref{fig5}, $B_1^{+}$ level created by the patch array is 5.3 times higher than for a stripline array at the same applied power. Note that due to the shape of the phantom ($x$ size of the phantom is substantially smaller than $y$ size), for both arrays, the dominant contribution to $B_1^{+}$ in the center of the phantom is from the $x$ component of the magnetic field. 

\begin{figure}
\center
\begin{minipage}{1\linewidth}
\center{\includegraphics[width=0.85\linewidth]{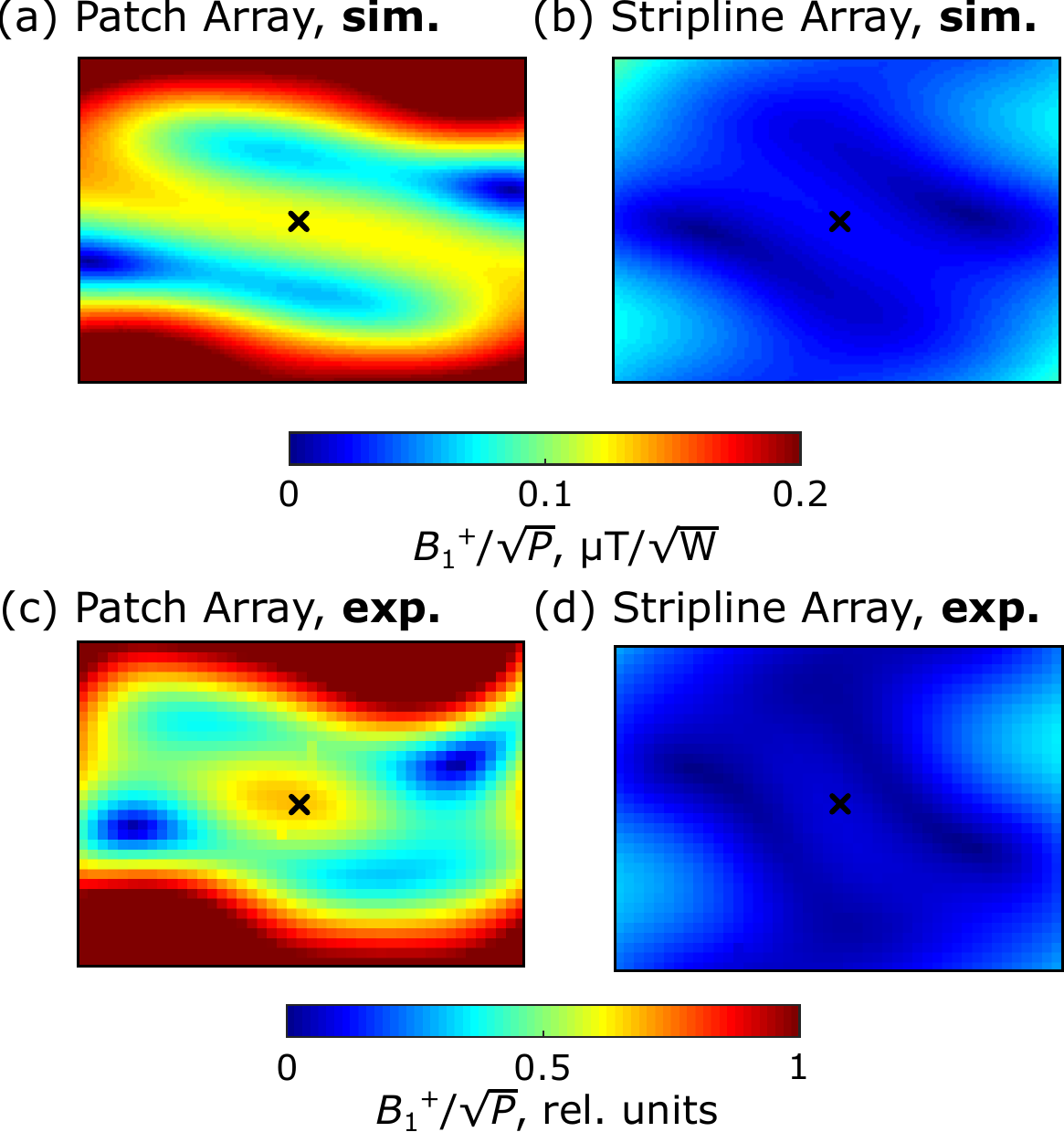}}
\end{minipage}
\caption{Numerically calculated (a,b) and experimentally measured (c,d) distributions of Tx efficiency field in the rectangular region of the central transverse plane of the homogeneous phantom for the patch (a,c) and stripline (b,d) arrays. The crosses show the center of the phantom.}
\label{fig6}
\end{figure}

Next, the experimental comparison results are presented. The experimental array of patch antennas in the presence of the phantom has the following measured characteristics. The median coupling $|S_{i,j}|$ between various pairs of eight elements $(i\ne j)$ is -10.2 dB with the worst value of -7.8 dB (for the two nearest patch antennas). The reflection coefficient level ranges from -22 to -7 dB, with the most inputs matching better than those with a level of -10 dB. The elements of the stripline array were all matched with a reflection coefficient level of better than -10 dB and a median value of -19.5 dB. The inter-element coupling coefficient for each pair of matched striplines (without any decoupling methods applied) was lower than -20 dB.

Measurements of $B_1^{+}$ distributions are made with a magnetic-field probe positioned within the area in the central transverse plane of the phantom limited with the rectangle shown in Fig.~5 (with dimensions of $110\times 80$ mm$^2$). The corresponding cropped numerically calculated maps of both arrays are compared with the measured distributions in Fig.~\ref{fig6}. The shapes of the measured field distributions are in qualitative agreement with the calculated ones. The measured field levels for both arrays are normalized by the same value and are presented in relative units. However, since proper matching at antenna inputs of both arrays guarantees the exact total applied power during measurements, the values in Fig.~\ref{fig6}(c) (for the patch array) and Fig.~\ref{fig6}(d) for the stripline one can be directly compared together. For both arrays, a local maximum (lobe) at the center of the phantom is indicated with a cross, in which the patch array achieves 8 times higher $B_1^{+}$ than the stripline array.

\subsection{The comparison of eight-channel arrays in the presence of a voxel model of the human body}

Although the efficiency of both compared arrays can be estimated in the presence of a homogeneous phantom mimicking the human body, a more detailed numerical study can be made via simulation with inhomogeneous human body models. Using this method, one can verify the transmit efficiency improvement and predict peak SAR values in local hotspots.

\begin{figure}
\center
\begin{minipage}{1\linewidth}
\center{\includegraphics[width=0.85\linewidth]{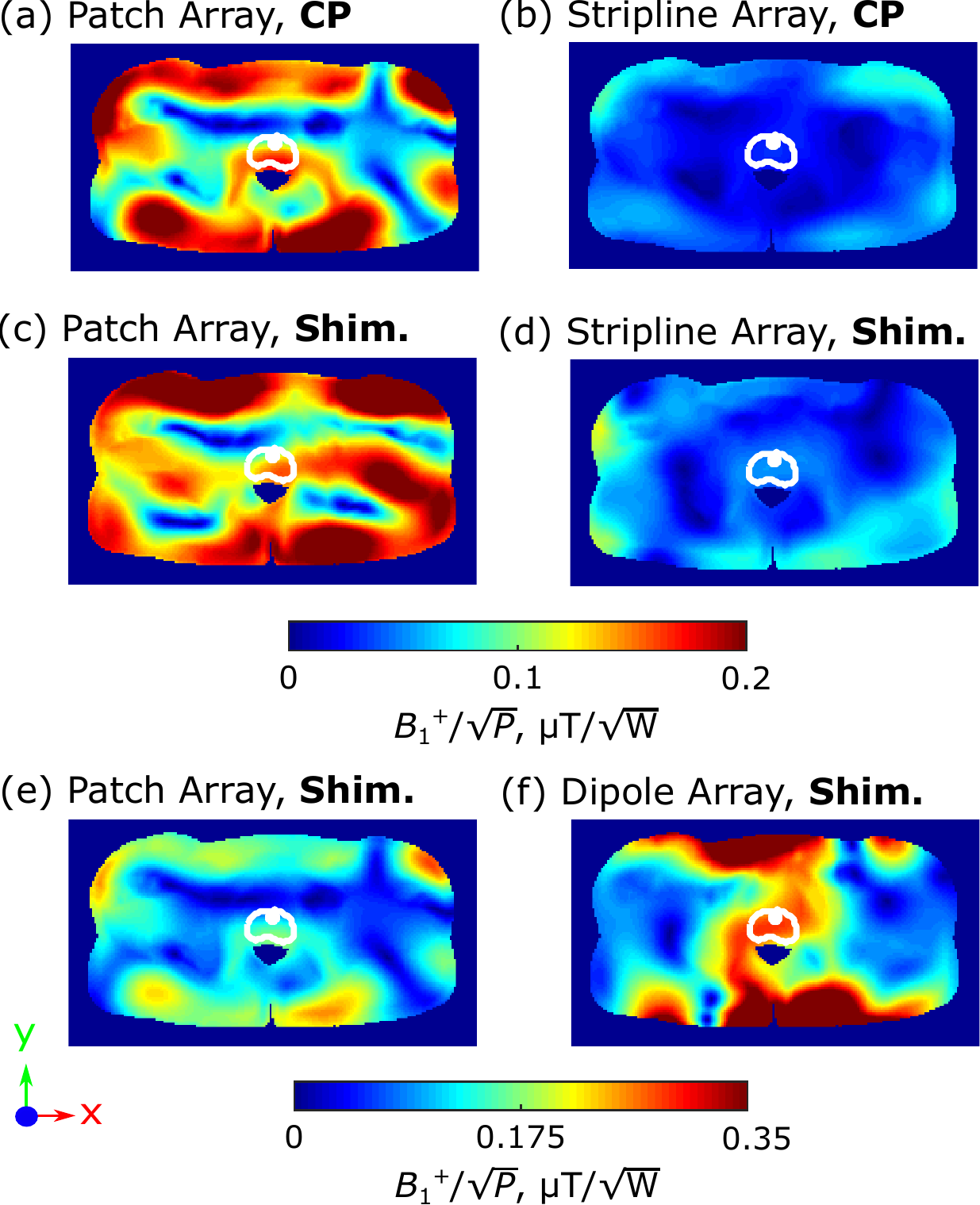}}
\end{minipage}
\caption{Numerically calculated distributions of Tx efficiency in the transverse plane through the prostate of Duke voxel human-body model for the patch (a,c) and stripline (b,d) bore-integrated arrays. The results are presented for the CP mode (a,b) and in the phase shimming regime with optimized phases maximizing the $B_1^{+}$ level in the prostate region (c,d). The same distribution as in (c) is given in (e) in a different color scale and compared with the corresponding distribution of the local fractionated dipole array in the phase shimming regime (f). The white contours show the cross-section of the prostate.}
\label{fig7}
\end{figure}

The numerically calculated $B_1^+$ distributions in the transverse plane of the human body model containing a cross-section of the prostate are presented in Fig.~7 for three different arrays: bore-integrated patch array, bore-integrated stripline array, and local (on-body) fractionated dipole array \cite{raaijmakers2016fractionated}. The two bore-integrated arrays are compared for two different excitation regimes, i.e., in the CP mode providing better homogeneity (a,b), and in the phase, shimming regime with optimized phases maximizing the $B_1^{+}$ level in the prostate region (c,d). For both bore-integrated arrays, the transmit efficiency differs only within 20--55\% for the considered regimes. However, the transmit efficiency of the dipole array in the prostate region drastically depends on the excitation regime due to the proximity of the antenna elements to the body model. Therefore, for the efficiency comparison, only the phase-only RF shimming \cite{metzger2008local} regime is considered for the dipole array (see the comparison with the patch array in Fig.~\ref{fig7}(e,f)). The calculated averaged transmit efficiency defined as $\langle B_1^{+}\rangle/\sqrt{P}$ with $\langle B_1^{+}\rangle/$ being the $B_1^{+}$ field level averaged over the prostate and $P$ is the total accepted power to the antenna array, the peak value of the SAR level averaged over 10-g volumes of body tissues (pSAR$_{\text{10g}}$), as well as the SAR efficiency defined as $<B_1^+>/\sqrt{\text{pSAR}_{\text{10g}}}$ for the considered arrays and excitation regimes are summarized in Table~I. 
\begin{table}[]

\center
\scriptsize
\begin{tabular}{|c|c|c|c|c|}
\hline
Array & Excitation & $\frac{\langle B_1^{+}\rangle/}{\sqrt{P}}$, $\frac{\mu \text{T}}{\sqrt{\text{W}}}$ & pSAR, $\frac{\text{W}}{\text{kg}}$ & $\frac{\langle B_1^{+}\rangle/}{\sqrt{\textnormal{pSAR}}}$, $\frac{\mu \text{T}}{\sqrt{\text{W/kg}}}$ \\ \hline
Frac. dip. & Phase shim.     & 0.26                                   & 0.67           & 0.32   \\ \hline
Patch       & Phase shim.     & 0.15                                  & 0.44           & 0.22   \\ \hline
Patch       & CP mode   & 0.12                                  & 0.18          & 0.28   \\ \hline
Strip.    & Phase shim.     & 0.05                                  & 0.18            & 0.12   \\ \hline
Strip.    & CP mode   & 0.03                                  & 0.08            & 0.11   \\ \hline
\end{tabular}
\caption{Numerically calculated Tx and SAR efficiency in the prostate region of the Duke multi-tissue model of the human body for different compared arrays and excitation regimes}
\label{tab1}
\end{table}

The proposed patch array provides 3.9 times higher averaged transmit efficiency in the CP mode (compared with the above-presented value of 5.3 for the homogeneous phantom) and 3.0 times higher efficiency for the phase shimming regime. pSAR$_{\text{10g}}$ value for the patch array is 2.2 times higher in CP mode and 2.5 times higher for the phase-shimming. Nevertheless, the SAR efficiency of the patch array is still 2.5 times higher in CP mode and 1.9 times higher than for a stripline array. The local array of fractionated dipoles exhibits 1.8 times higher averaged transmit efficiency efficiency and 1.5 times higher SAR efficiency versus the proposed array. The corresponding peak local SAR for the dipole array is also increased by 1.5 times.

\section{DISCUSSION}

The presented novel design of bore-integrated array based on patch antennas provides improved Tx efficiency compared to the \textit{state-of-the-art} bore integrated array based on the striplines\cite{orzada2017analysis}. The patch antenna element utilizes an RF shield as a part of the antenna array. Therefore, as shown by single element simulation, the patch is capable of creating a higher $B_1^+$-field both at the surface and in the depth of the homogeneous phantom (Figure \ref{fig1} (c-f)). However, we found a strong coupling between array elements when adding adjacent antennas. It is worth mentioning that coupling appears both because of the complex octagonal-shaped bore and the patch's reduced width. We used a magnetic resonator between the non-radiating edges of patch antennas close to the octagonal prism vertex to reduce coupling. Such resonator geometry was chosen to provide coupling to the symmetric and anti-symmetric modes of coupled patch antennas. This allowed us to create a fully decoupled eight-channel array. Numerical simulations of eight channel patch and stripline arrays for the CP mode excitation showed a significant increase of $B_1^+$ level inside the phantom. These results were proven by the experimental measurements using a near-field scanner. However, since the magnetic field probe could not measure the absolute value, only relative values of $B_1^+$ were measured. Also, we could not reach the exact value of input matching and decoupling between adjacent elements for the patch array. Lower decoupling levels in experiments were not obtained because of the instability of the position of the decoupling resonator and the necessary usage of the eight-port network analyzer to adjust simultaneously the parameters of all elements. The patch array showed a 5.3 times higher Tx efficiency in the numerical simulations and an 8.0 times higher Tx efficiency in measurements compared to the stripline array in the center of the phantom.

Numerical simulations of the array using human body multi-tissue also showed improved Tx efficiency compared to the bore-integrated stripline array. However, the patch array's peak SAR$_{10g}$ was also higher. This corresponds to the general rule of thumb of high-field MRI RF coil design: higher Tx efficiency requires higher tissue power deposition, which leads to higher SAR. However, the SAR efficiency of the proposed array was more than two times higher for the phase-only shimming and CP mode excitation. 

\section{CONCLUSION}

In this work, we developed, constructed, and evaluated numerically and experimentally an eight-element bore-integrated patch array for 7T body imaging. In comparison to the bore-integrated stripline array with the same size and number of elements, the proposed novel design grants drastically higher transmit and SAR efficiencies in the presence of a homogeneous phantom (5.3 times in the numerical simulations and 8.0 times in experiments) and also in the numerical simulations in the presence of a multi-tissue voxel model (3.0--3.9 times depending on the excitation regime). This considerable benefit of the proposed antenna element results from much lower internal dissipation losses (83\% vs. 11\% for the stripline in the single channel setup). The proposed design can be tuned and matched like a conventional patch antenna using a series capacitor and selecting a proper feed position. As a drawback, the patch array for 7T MRI requires additional decoupling elements to be installed in the gaps between adjacent elements. Such decoupling resonators were designed and described in this work. Despite the proposed array not reaching the efficiency levels of the tight-fit local dipole array, it approached those levels quite well (0.78 of Tx efficiency of fractionated dipole array in the phase shimming regime). At the same time, being more efficient, the proposed array provides the same free space for a human body space as the previously demonstrated bore-integrated array.

In conclusion, the novel patch array opens a new direction in developing highly efficient bore-integrated arrays for body imaging. We believe this design is worth further adaptation and characterization in frames of a research 7T MR system. Note that such experimental investigations require a substantial modification of the MR system similar to those performed in works \cite{orzada2017analysis,orzada201932}. Also, more complex RF shimming strategies will be investigated using the proposed patch array. Future work, including $B_1^{+}$ and SAR mapping using MR techniques on phantom and \textit{in-vivo} trials made for the transmitting patch array in conjunction with a local receive coil, should improve the state-of-the-art in the UHF MRI.

\section{ACKNOWLEDGMENTS}
Numerical studies are supported by the Ministry of Science and Higher
Education of the Russian Federation (Project No. 075-15-2022-1120). Experimental studies are supported by the Russian Science Foundation (Project 21-19-00707). We also gratefully acknowledge Alexander Kalganov and Alexandra Dudnikova's help with the experimental setup and near-field measurements.

\bibliographystyle{IEEEtran}
\bibliography{references}

\end{document}